\documentclass[journal,transmag]{IEEEtran}
\usepackage{graphicx}
\usepackage{lipsum}
\usepackage{xcolor}
\usepackage{threeparttable}
\usepackage{enumitem}
\usepackage{chngcntr}
\usepackage{graphicx}
\usepackage{textcomp}
\usepackage{xcolor}
\usepackage{multirow}
\usepackage{multicol}
\usepackage{CJKutf8}
\usepackage{booktabs} 
\usepackage{multirow}
\usepackage{amsmath}
\usepackage{amssymb}
\usepackage{nicefrac}
\usepackage{booktabs}  
\usepackage{tabularx}
\usepackage{geometry}
\usepackage{graphicx}
\usepackage{subcaption}
\usepackage{hyperref}

\usepackage{cite}

\ifCLASSINFOpdf
\else
\fi
\usepackage{amsmath,amssymb,amsfonts}
\usepackage[linesnumbered, ruled]{algorithm2e}

\usepackage{array}
\def\BibTeX{{\rm B\kern-.05em{\sc i\kern-.025em b}\kern-.08em
    T\kern-.1667em\lower.7ex\hbox{E}\kern-.125emX}}

\begin{document}
%
\title{BIDO: An Out-Of-Distribution Resistant Image-based Malware Detector}


\author{\IEEEauthorblockN{Wei Wang\IEEEauthorrefmark{1}, Junhui Li\IEEEauthorrefmark{1}, Chengbin Feng\IEEEauthorrefmark{2}, Zhiwei Yang\IEEEauthorrefmark{1}
and Qi Mo\IEEEauthorrefmark{1}}
\IEEEauthorblockA{\IEEEauthorrefmark{1}Software School,
Yunnan University, Kunming, Yunnan 650091 China}
\IEEEauthorblockA{\IEEEauthorrefmark{2}School of Information Systems, University of New South Wales, Sydney, NSW 2052 Australia}
\thanks{Manuscript received December 1, 2012; revised August 26, 2015. 
Corresponding author: Qi Mo (email: moqi@ynu.edu.cn).}}

\markboth{Journal of \LaTeX\ Class Files,~Vol.~14, No.~8, August~2015}%
{Shell \MakeLowercase{\textit{et al.}}: Bare Demo of IEEEtran.cls for IEEE Transactions on Magnetics Journals}
%



\IEEEtitleabstractindextext{%
\begin{abstract}
While image-based detectors have shown promise in Android malware detection, they often struggle to maintain their performance and interpretability when encountering out-of-distribution (OOD) samples. Specifically, OOD samples generated by code obfuscation and concept drift exhibit distributions that significantly deviate from the detector's training data. Such shifts not only severely undermine the generalisation of detectors to OOD samples but also compromise the reliability of their associated interpretations. To address these challenges, we propose BIDO, a novel generative classifier that reformulates malware detection as a likelihood estimation task. 
Unlike conventional discriminative methods, BIDO jointly produces classification results and interpretations by explicitly modeling class-conditional distributions, thereby resolving the long-standing separation between detection and explanation.
Empirical results demonstrate that BIDO substantially enhances robustness against extreme obfuscation and concept drift while achieving reliable interpretation without sacrificing performance. The source code is available at \url{https://github.com/whatishope/BIDO/}.
\end{abstract}

\begin{IEEEkeywords}
Android Malware Detection, APK Image, Generative Classification, Trustworthiness.
\end{IEEEkeywords}}

\maketitle

\IEEEdisplaynontitleabstractindextext

%
\IEEEpeerreviewmaketitle

\section{Introduction}
%
%
%
%
\IEEEPARstart{D}{ue} to their versatility, Android applications (apps) have deeply transformed our daily lives, serving as an enabling technology across various disciplines. Despite these rapid advances, malicious apps (malware) designed to harm systems, networks, or users have become a significant concern. According to \cite{avtest}, the number of malware increased to 22,184,323 in 2022, approximately 2405 times compared ten years ago. Accordingly,  malware detection has attracted considerable attention. 

Although various malware detection techniques have been introduced in the past decade, image-based detectors have become prominent primarily due to their efficiency, as they completely eliminate the need for reverse engineering processes, such as disassembly, and specialised execution environments, like sandboxes, required by conventional static and dynamic detection methods \cite{qiu2020survey}. Moreover, by formulating malware detection as an image classification  problem, state-of-the-art deep learning (DL) algorithms (e.g., CNN and GNN) can be seamlessly integrated into these image-based detectors, thereby ensuring the  scalability  required  to  process  massive samples in modern malware detection \cite{moawad2022survey}.

Despite the aforementioned advantages, image-based detectors are mainly based on discriminative DL algorithms. When faced with pervasive out-of-distribution (OOD) instances introduced by code obfuscation \cite{Hammad_Garcia_Malek_2018} and concept drift \cite{pendlebury2019tesseract}, image-based detectors typically suffer from two drawbacks. First, robustness issue. By distorting feature space, code obfuscation and concept drift enforce OOD instances misaligned with the pre-defined decision boundaries, leading detectors cannot adapt to the changing distribution. Second, unreliable interpretation. Most interpretation approaches for image-based maware detectors follow post-hoc paradigm, where interpretation is decoupled from model training. When these methods are applied to OOD samples, the discrepancy between the training data distribution and the OOD data distribution undermines the validity of the generated interpretations. As a result, the interpretations fail to faithfully reveal the model's underlying decision logic.

To address these challenges, we propose BIDO, a novel image-based malware detector. Unlike existing solutions\cite{kan2021investigating, fernando2022fesa, haq2022maldroid} that treat concept drift and code obfuscation independently, BIDO addresses both issues within a unified generative framework. Specifically, BIDO extracts configurational and functional features from malware images and encodes their cross-modal dependencies within an Outer Product Space (OPS). These OPS representations are then mapped into a Mixture Gaussian Distribution (MGD) via the revised Normalizing Flow. The class-conditional likelihood of each input serves as the foundation for classification, while its decision logic is explicitly interpreted through the distances between the latent representation of the input and the centroids of the MGD. Furthermore, the confidence of the interpretation is quantitatively measured by the likelihood itself.

Compared with existing methods, our core contributions are as follows:
\begin{itemize}
\item \textbf{Generative Classification Framework.} We propose a generative malware classifier that represents apps in a probabilistic space rather than a deterministic vector space.  Each app in this space is associated with a Gaussian distribution and a class-conditional likelihood. By explicitly modeling the uncertainty of individual samples, this probabilistic representation mitigates the sensitivity to distributional shift and enables more robust decision-making under OOD scenarios. To the best of our knowledge, BIDO is the first image-based detector that can effectively generalize to OOD cases.

\item \textbf{Synergistic Detection and Interpretation.} Rather than treating malware detection and interpretation as two isolated stages, BIDO explicitly integrates interpretability into the detection framework as a core design objective. In this unified paradigm, detection and interpretation are mutually reinforced: accurate detection yields a reliable interpretation, while reliable interpretation provides explicit guidance for further optimizing the detection process.

\item \textbf{Extensive Empirical Validation.} Through comprehensive experiments, we demonstrate that generative modeling is a highly effective paradigm for simultaneously improving the robustness and interpretability of image-based malware detectors against OOD threats.
\end{itemize}

The rest of the paper is organized as follows: Section 2 discusses related work. Section 3 presents an example to illustreate the motivation of this paper. Section 4 presents the details of our approach. Section 5 presents the experimental results. Section 6 summarizes the work presented in this paper.

\section{Related Work}
This section provides an overview of recent advances in malware detection, highlights the limitations of existing approaches and articulates the research motivations behind our work.

\subsection{Malware Detection}
Existing malware detectors are broadly categorized into static and dynamic methods \cite{qiu2020survey}. Despite their high accuracy, dynamic methods suffer from limited scalability due to their heavy reliance on complex, resource-intensive simulation environments, such as sandboxes \cite{alzaylaee2020dl}. Consequently, the majority of research has shifted toward static analysis \cite{gao2024comprehensive}.
Traditional static methods employ diverse reverse-engineering tools (e.g., Apktool\footnote{https://apktool.org/}, Androguard\footnote{https://github.com/androguard/androguard}) to extract opcode sequences, API permissions,  control/data-flow graphs and other features from configuration (\texttt{AndroidManifest.xml}) and DEX (\texttt{classes.dex}) files. Advanced deep learning architectures, including RNNs \cite{lakshmanarao2022android}, LSTMs \cite{shen2023self}, Graph Neural Networks (GNNs) \cite{yumlembam2022iot}, and node2vec \cite{cui2023api2vec} are subsequently utilized to encode these features into embeddings for classification. However, the application of these methods is greatly weakened in practice, as the requirement of complex and fragile reverse engineering infrastructures \cite{qiu2020survey}.

In contrast, image-based methods offer an effective alternative by eliminating reverse engineering entirely. Both APK files and images consist of hexadecimal data. This technical similarity makes it feasible to directly convert APKs into grayscale or RGB images \cite{vasan2020image, mercaldo2020deep}, thereby mitigating the need for reverse-engineering tools. A key advantage of this approach is that it transforms malware detection into a image classification task, facilitating the seamless integration of advanced DL algorithms while significantly expanding application scope \cite{daoudi2021dexray}. Following the pioneering work of Nataraj et al. \cite{nataraj2011malware}, numerous novel detectors have been proposed in recent years. For instance, Xiao et al. \cite{Xiao_Yang_2019} transformed DEX files into RGB images and proposed a CNN-based detector. Daoudi et al. \cite{daoudi2021dexray} generated grayscale “vector” images from DEX files and applied a 1D CNN for malware detection.

\subsection{OOD in Malware Detection}
OOD samples refer to those whose underlying distribution differs from that of the training data. In the context of malware detection, code obfuscation and concept drift are two main sources of OOD instances \cite{gao2024comprehensive}. By modifying the bytecode structure or semantics of APKs, code obfuscation causes APK images to exhibit visual patterns that are distinct from their original forms \cite{Garcia_Hammad_Malek_2017}. Thus, results in significant performance degradation \cite{gao2024comprehensive, zhu2023effective}. 
Furthermore, concept drift introduces distribution shifts through the continuous evolution of apps \cite{Wang_Wang_Zhan_Wang_Liu_Luo_Cheung_Wang}. Singh \textit{et al.} \cite{singh2012tracking} reported that concept drift degraded the performance of detectors by (1) the continuous changes in their structural and semantic patterns, and (2) new categories of malware emerge. 

Existing detection methods treated code obfuscation and concept drift as distinct challenges and employ advanced DL algorithms \cite{garcia2023effectiveness, hu2017concept} or robust feature selection techniques (such as SIFT, SURF, and KAZE descriptors \cite{unver2020android} or max pooling \cite{zhu2023effective}) to enhance the robustness of representations \cite{fernando2022fesa}. These methods assume that coarse-grained representations are more resistant to code obfuscation or concept drift \cite{gao2024comprehensive}.  However, existing solutions follows the discriminative paradigm that typically requires the classisication probabilities must sum to 1. This constraint forces the detectors to assign high probabilities to OOD instances, even if they do not resemble any known malware classes \cite{brosolo2025road, mackowiak2021generative}.


\subsection{Interpretation to Malware Detection}
Interpretability refers to the degree to which a human can understand a system's decision-making logic. Reliable interpretation is essential for ensuring the trustworthiness of detection results while enabling broader application scope \cite{moawad2022survey}. Efforts in this area can be divided into global and local strategies \cite{saqib2024comprehensive}. 
Global strategies aim to embed interpretability directly into the model architecture (e.g., decision trees or rule-based systems) to provide an overview of the model's behavior across the entire dataset.
However, these methods often struggle to offer detailed interpretations for individual inputs which are critical for generating actionalbe guidance for future activities \cite{saqib2024comprehensive}. As a result, many practical applications have adopted local methods instead. 

Local strategies \cite{ribeiro2016should, lundberg2017unified} focus on explaining individual predictions. Most of them follow the post-hoc paradim, where the detector is first trained and explanations are generated subsequently by XAI techniques such as LIME \cite{ribeiro2016should} or BreakDown \cite{gosiewska2019not}. Despite their widespread adoption, these methods often yield unreliable interpretations, particularly in OOD scenarios \cite{xu2025joint}.
Because OOD samples are absent from the training distribution, the detector fails to learn meaningful representations for them. Consequently, its decision-making on such data is frequently arbitrary or driven by spurious correlations. As a result, post-hoc explanations in these contexts fail to reflect any semantically grounded reasoning process \cite{alvarez2018robustness}.

\subsection{Summary}
Based on the literature review above, existing studies on malware detection typically focus on static methods. Among these, image-based techniques have gained significant attention due to their superior computational efficiency compared to other static alternatives. However, most existing image-based methods rely on discriminative models, which are susceptible to OOD samples. Due to the inability to learn the intrinsic representations of the OOD samples, their decision boundaries become ill-defined. Consequently, this issue directly undermines the reliability of post-hoc interpretations, as they attempt to reveal unreliable decision processes for OOD samples.

\section{Motivating Example}

\begin{figure}[tb]
    \centering
    \begin{subfigure}[t]{0.5\textwidth}
        \centering
        \includegraphics[width=\textwidth]{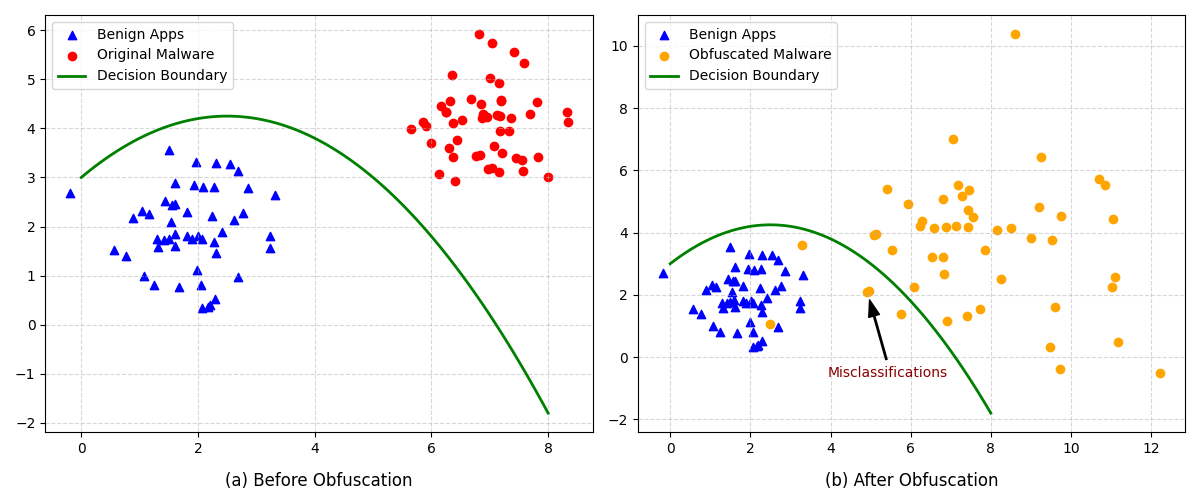} 
        \label{fig:obfuscation}
    \end{subfigure}   
    \begin{subfigure}[t]{0.5\textwidth}
        \centering
        \includegraphics[width=\textwidth]{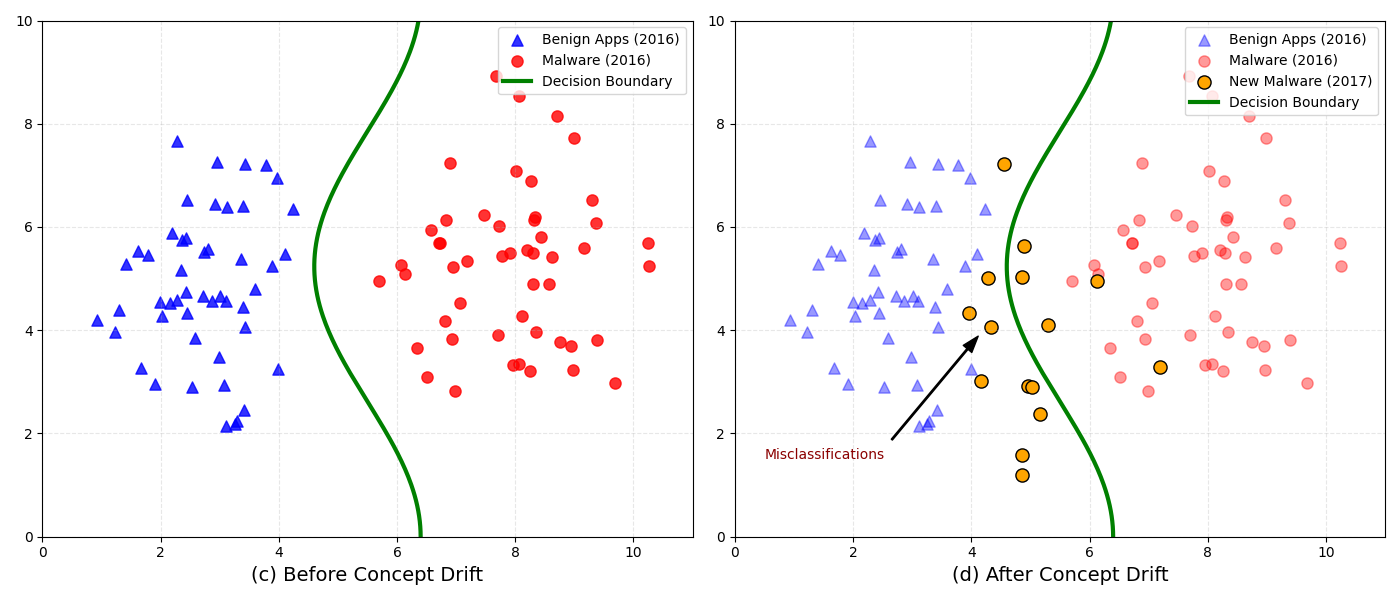} 
        \label{fig:concept_drift}
    \end{subfigure} 
    \caption{Impacts of OOD samples to Malware Detection}
    \label{fig:combined_analysis}
\end{figure}

First, to illustrate the impact of code obfuscation to the malware detection, we constructed a dataset comprising 100 samples: 50 benign apps and 50 malware samples from \textit{Swizzor.gen!E} family. To visualize these samples, we projected the high-dimensional APK images into a two-dimensional space. As shown in Fig. \ref{fig:combined_analysis}(a), blue triangles represent benign apps, while red circles denote malware samples. We adopt the MADRF-CNN \cite{zhu2023effective} a recently proposed detector as a classifier on this dataset. As presented in FIG. \ref{fig:combined_analysis}(a), the decision boundary (indicated by the green line) can accurately separate benign apps from malicious ones. Then we implement three obfuscations, Rename, ResStringEncryption and Control Flow Obfuscation to malware samples. As shown in Fig. \ref{fig:combined_analysis}(b), obfuscations result in a radically distribution shift: the obfuscated \textit{Swizzor.gen!E} samples (orange circles) diverge from their original cluster. Consequently, the pre-established classification boundary fails to generalize well, leaving many obfuscated samples on the "benign" side of the boundary.

Second, to illustrate the impact of concept drift to the malware detection, we constructed another dataset of 100 samples from 2016: 50 benign apps and 50 malware samples. Similar to the last example, blue triangles and red circles denotes benign and malware samples. The green line is the decsion boundary. Then we add 15 malware samples in 2017 (orange points) into the dataset. While these new samples share the same malicious intent, their feature distributions deviate from the training set. As shown in Fig. \ref{fig:combined_analysis}(d), the decision boundary fails to adapt to this shift, resulting in many orange points in misclassifications.

Moreover, OOD samples undermine the reliability of interpretation in malware detection. As illustrated in Fig. \ref{fig:combined_analysis}, when the classifier is trained on an ordinary dataset, its decision boundary is shaped by meaningful features that consistently differentiate benign and malicious samples. Consequently, interpretations such as feature attribution can reliably trace a prediction back to those discriminative features that actively contribute to crossing the decision boundary, thereby providing a faithful explanation of the detection result. However, code obfuscation and concept drift introduce samples that were not observed or validated during training, resulting in the attributed features to no longer correspond to the actual reasons behind the detection result. As a result, interpretations outputs no longer reflect the actual decision logics of detection.

To enhance the robustness of malware detection and the reliability of interpretation, we propose BIDO, a novel generative model that redefines malware classification as a likelihood estimation task and integrates detection and interpretation into a single framework. 

\section{Our Approach}

\begin{figure*}[t]
	\centering
	\includegraphics[width=0.96\linewidth]{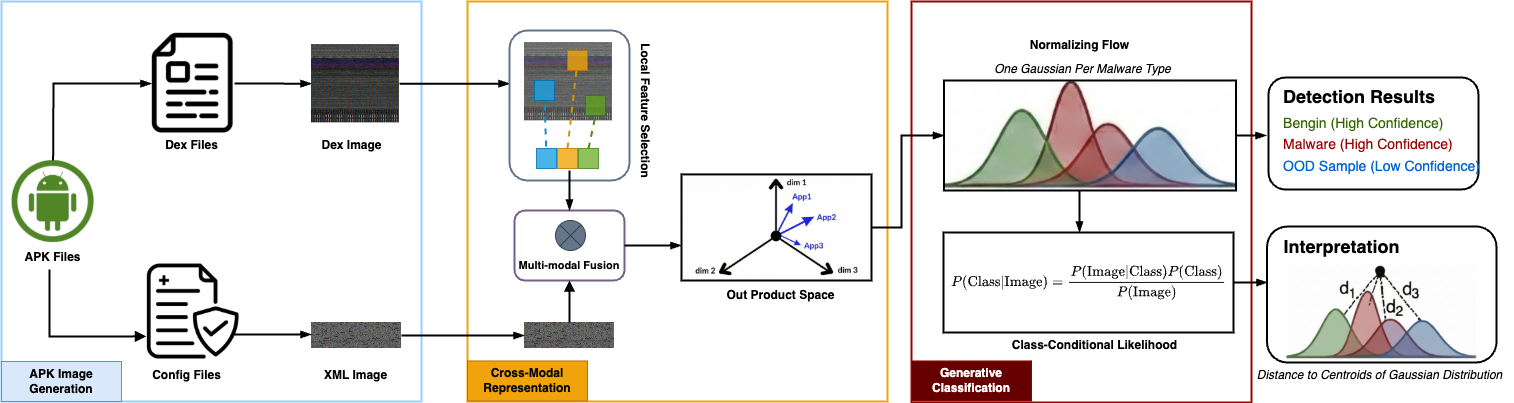}
	\caption{The Architecture of Our Method}
	\label{architecture}
\end{figure*}

As shown in Fig. \ref{architecture}, the BIDO consists of three modules: (1) the APK image generation module, which converts configuration and DEX files of APKs into images; (2) the cross-modal representation module, which represents the configuration and DEX images in a shared low-dimensional space; and (3) the generative classification module, which transforms the low-dimensional embeddings into a mixture Gaussian distribution and then generates detection and interpretation results.

\subsection{APK Image Generation}
Different from existing methods, we only convert the index part of the DEX files into images, discarding the header and data parts. The reasons for our decision are as follows:

\begin{itemize}
    \item The header part primarily defines the offsets of other parts which does not contain any semantic information relevant to malware detection.
    \item The data part constitutes approximately 80\% of a DEX file. Even in malware, the majority of this part comprises benign elements. Besides increaes the size of DEX image, converting it into the image may also increase the tendency of blindly classifying apps as benign.
    \item The index part is consisted of the string index, the type index, the proto index, the field index, and the method index. These elements directly reflect what the app can do. It is highly correlated with malicious behavior.
\end{itemize}

We adopt the algorithm proposed in \cite{zhu2023effective} to transform the index section into an image. Specifically, every six hexadecimal numbers are first converted into three decimal digits through “and" and “shift" operations and then form a three-channel pixel. For instance, the hexadecimal digits \texttt{0x868812} corresponds to the pixel values (R=134, G=136, B=18). To ensure consistent dimensions, images are padded with zeros as needed. Ultimately, the DEX image is represented as  $I_{dex}\in\mathbb{R}^{H_d\times W_d\times 3}$, where $H_d$ and $W_d$ are the height and width of the feature map. Additionally, the  \texttt{AndroidManifest.xml} file is also hexadecimal file and we also convert it into an image with the similar process of DEX-image generation. The configuration image is presented as $I_{xml}\in \mathbb{R}^{H_x\times W_x\times 3}$, where $H_x, W_x$ are the height and width of the image.

\subsection{Cross-Modal Representation}
The primary challenge in directly applying generative learning techniques to the APK image classification is the huge dimension of APK images \cite{Fetaya2019ConditionalGM, fetaya2019understanding}. 
Generative models aim to learn the data distribution $p(\mathrm{Image}|\mathrm{Class})$ or $p(\mathrm{Image})$, which requires modeling complex dependencies among pixels. 
Unlike natural images, where local spatial patterns correspond to meaningful visual concepts, APK images encode program bytes whose spatial adjacency does not necessarily reflect semantic relationships in code. 
As dimensionality increases, the number of samples required to accurately estimate the distribution grows exponentially, making the learned likelihood poorly calibrated.

\begin{figure}[t]
	\centering
	\includegraphics[width=0.96\linewidth]{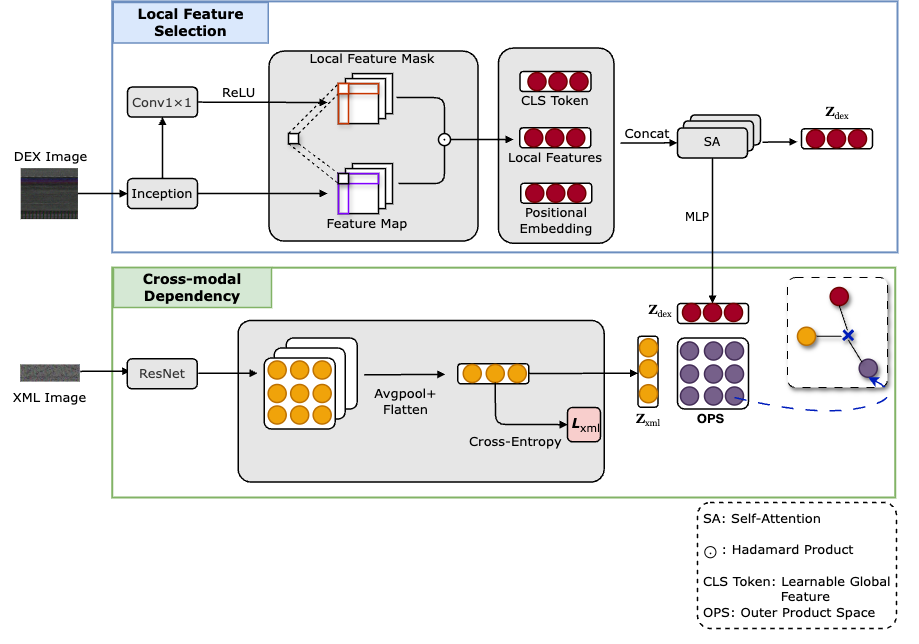}
	\caption{The Process of Corss-modal Representation.}
	\label{OPS}
\end{figure}

To solve this issue, 
as shown in Fig.~\ref{OPS}, we represent the configuration image and the DEX image in an outer product space (OPS) that not only preserves discriminative information but also meet the low-dimensional requirement of subsequent generative classification.

\subsubsection{Local Feature Selection}
To identify highly informative byte-code patterns, such as suspicious API usage or specific opcode sequences, we propose the local feature selection module. Specifically, a pretrained network is first used to extract feature maps $F_{dex} \in \mathbb{R}^{ H_d \times W_d\times C}$ from DEX-images, where $C$ is the number of channels. Then we apply a $1\times 1$ convolution kernel $\phi\in \mathbb{R}^{1\times1\times C}$ to these feature maps to generate feature mask matrix $M \in \mathbb{R}^{ H_d \times W_d}$. Each element of $M[i,j]$ indicates if the corresponding pixel in $F_{dex}$ is informative.
Given a feature map  $F_{dex}$ and its learned masks $\{M_1, M_2, \cdots, M_k\}$, the local feature map set  $L_f=\{L_f^1, \cdots, L_f^k\}$ is defined as $L_f^i=F_{dex}\odot M_i$, where $\odot$ is the Hadamard product.

Furthermore, unlike natural images, Dex images encode serialized program structures, where semantically related components (e.g., String index and Method index) may be spatially distant.
To aggregate all contextual information together, we employ a self-attention mechanism in this paper. Given the local feature set $L_f\in \mathbb{R}^{H_d\times W_d \times k}$, we compute the contextual representations as:
\begin{equation}
    E=\mathrm{SoftMax}\left(\frac{Q_fK_f^\top}{\sqrt{d}}\right)\odot V_f
\end{equation}
where $Q_f=X_fW_Q,K_f=X_fW_K,V_f=X_fW_V$ and $W_Q$, $W_K$, $W_V$ and $CLS\in \mathbb{R}^{H_d\times W_d \times 1}$ are learnable matrices. $X_f=[L_f; CLS]+P$ and $P\in\mathbb{R}^{H_d\times W_d\times (k+1)}$ is positional embedding. $d$ is the length of the local feature map and $\mathrm{SoftMax}$ refers to the softmax normalization.

\subsubsection{Cross-modal Dependency Representation}
In this section, we project the DEX-image and XML-image into an OPS. Given the contextual representations of the DEX-image $E$, and the XML-image $I_{xml}$, we first transform them into two latent vectors, $Z_{dex} \in \mathbb{R}^l$ and $Z_{xml} \in \mathbb{R}^h$, via MLPs. Subsequently, we formulate the cross-modal dependency matrix $D \in \mathbb{R}^{l \times h}$ as follows:
\begin{equation}
D=\mathrm{NOR}(Z_{xml}\otimes Z_{dex})
\end{equation}
where $\otimes$ denotes the outer product operation and $\mathrm{NOR}(\cdot)$ represents a normalization function. Each element $D[i,j]=Z_{xml}[i]\cdot Z_{dex}[j]$ in the OPS captures the similarity between $Z_{xml}[i]$ and $Z_{dex}[j]$. A higher similarity indicates a stronger correlation between the corresponding configuration and DEX features. The normalization technique ensures the subsequent malware classification module focuses on the pattern of the feature rather than its magnitude scale.

Comparing with conventional information fusion techniques, such as self-attention, element-wise addition or feature concatenation, the OPS offers the following benefits:
\begin{itemize}
    \item \textbf{Enhanced robustness}. OPS captures cross-modal co-occurrence patterns, which are more robust than features derived from a single modality. These persistent correlations ensure the representation remains robust even if individual modalities are compromised by obfuscation or concept drift.
    \item \textbf{Information compensation}. OPS uncovers subtle semantic connections by encoding exhaustive pairwise dependencies. This capability allows the model to extract hidden signals, effectively compensating for the information loss caused by code obfuscation or concept drift.
\end{itemize}

However, the OPS may enlarge the size of feature maps \cite{yu2021fast}. To address this issue, we utilize the factorization technique presented in \cite{gao2020revisiting} to reduce the size of feature maps while preserving the information of cross-modal dependencies. 

\subsection{Generative Classification}\label{GC}
The generative classification module comprises two components: the classifier and the loss function.
\subsubsection{Classifier Architecture}
In this section, we construct a generative classifier based on the Normalizing Flow \cite{rezende2015variational}, a class of likelihood-based generative models that learn exact and tractable data densities through a sequence of invertible transformations. Given the cross-modal representation $Z_{ops}$, our goal is to approximate its probability $p(Z_{ops})$ via a flow-based model $q_\theta$. $q_\theta$ is a bijective neural network that transforms the complex input distribution $p(Z_{ops})$ into a latent Gaussian distribution $p(Z_{lat})\sim \mathcal{N}(0,1)$. The whole transforming process is governed by the change-of-variables formula:
\begin{equation}
	q_\theta(Z_{ops}) = p\left(f_\theta(Z_{ops})\right) \left| \det \frac{\partial f_\theta(Z_{ops})}{\partial Z_{ops}} \right|
\end{equation}
where $J = \partial f_\theta / \partial Z_{ops}$ denotes the Jacobian matrix of the transformation. We adopt the affine coupling architecture proposed in \cite{dinh2016density} as $q_\theta(\cdot)$, which can not only be used to estimate likelihoods $q_\theta(x)$, but also can map instances sampled from $Z_{lat}$ back to cross-modal space $Z_{ops}$.

However, standard normalizing flow in our case is insufficient for the purpose of malware classification. It requires to find the class $y$ under which the representation $Z_{ops}$ has the highest likelihood $q_\theta\left(Z_{ops}|y\right)$. To bridge this gap, we extend the latent space $Z_{lat}$ to a Gaussian Mixture Model (GMM) rather than a unimodal Gaussian in \cite{rezende2015variational}. The conditional distribution in the latent space is defined as $p(Z_{lat}|y) = \mathcal{N}(Z_{lat}; \mu_y, \mathbf{I})$, where $\mu_y$ represents the class-specific centroid and $\mathbf{I}$ is the identity covariance matrix. The reasons for integrating the GMM is threefold:
\begin{itemize}
    \item \textbf{Classification Tractability.} The classification result can be easily computed based on the Bayesian rule:
\begin{equation}
    \begin{aligned}
        p(y|Z_{lat})&= p(y)p(Z_{lat}|y)/p(Z_{lat})\\
        &=p(y)\mathcal{N}(Z_{lat};\mu_y,\mathbf{1})/p(Z_{lat}) 
    \end{aligned}
\end{equation}
where class priors $p(y)$ is the frequency of each type of malware in the dataset. 
\item \textbf{Interpretability.}  The multi-centroid nature of the GMM aligns well with the diversity of malware subtypes. By mapping each class of malware to a distinct Gaussian distribution, the latent space becomes structured. The classification decision is interpreted by the distance between the latent representation of an input image $Z_{lat}$ to the surrounding classes centroids $\mu_y$. The confidence of the decision is quantified by the conditional likelihood $p(Z_{lat}|y)$. 
\item \textbf{OOD Detectability.}  If an input image is mapped to a latent point far from all known Gaussian distribution, this suggests it could be an OOD sample.
\end{itemize}

\subsubsection{Loss Function}
Relying solely on the negative log-likelihood of normalizing flows often yields suboptimal performance for classification tasks \cite{fetaya2019conditional}. To address this, we construct a hybrid loss function that ensure the latent representation of each malware type align well with a specific Gaussian mode, while OOD samples are assigned low likelihoods.

Specifically, the loss function is defined as:
\begin{equation}
    L=L_G+L_C
\end{equation}
Minimizing the first term enforce the latent Gaussian feature captures more useful information of input images, while minimizing the second term boost the classification performance. 
The first term is defined as a negative log likelihood:
\begin{equation}
\begin{aligned}
    L_{G}(x)&=-\sum_{k=1}^n \log q_{\theta_k}(x)\\
    &=-\log |\mathrm{det}J(x)|+\sum_{k=1}^n \log \exp\left(d_{k}^2-p\right)
\end{aligned}
\end{equation}
where $n$ is number of class. $p=\log (1/n)$ is the uniform class priors. $d_k=q_\theta(x)-\mu_k$ is the discrepancy between generated latent space and the predefined Gaussian distribution (denoted by $\mu_k $). $\mathrm{det}J(x)$ is the Jacobian matrix.  This loss enforces the mapped latent points $q_\theta(x)$ to follow the multimodal Gaussian distribution, effectively mitigating the mode collapse issue often seen in unimodal flows.

To optimize malware classification, the second term is defined as follows:
\begin{equation}
L_{C} = \alpha L_{xml} + \beta L_{dex} + \pi L_{lat}+\gamma L_{ops} + \delta L_{con}
\end{equation}
$L_{xml}, L_{dex}, L_{ops}, L_{lat}$ are used to evaluate the consistency of classification results generated by $Z_{xml}$, $Z_{dex}$, $Z_{ops}$ and $Z_{lat}$ to the ground truth. Hyperparameters $\alpha,\beta$, $\pi$, $\gamma$  and $\delta$ are used to quantify the importance to the total loss. Specifically, these loss functions are defined as follows,
\begin{equation}
    \begin{aligned}
        L_{xml}&=-\frac{1}{T}\sum_{i=1}^T y_i\log \hat{y}_{xml}^i\\
        \hat{y}_{xml}^i&=\mathrm{SoftMax}(\mathrm{FC}(Z_{xml}))
    \end{aligned}
\end{equation}
\begin{equation}
    \begin{aligned}
        L_{dex}&=-\frac{1}{T}\sum_{i=1}^T y_i\log \hat{y}_{dex}^i\\
        \hat{y}_{dex}^i&=\mathrm{SoftMax}(\mathrm{FC}(Z_{dex}))
    \end{aligned}
\end{equation}
\begin{equation}
    \begin{aligned}
        L_{lat}&=-\frac{1}{T}\sum_{i=1}^T y_i\log \hat{y}_{lat}^i\\
        \hat{y}_{lat}^i&=\mathrm{SoftMax}(\mathrm{FC}(Z_{lat}))
    \end{aligned}
\end{equation}
\begin{equation}
    \begin{aligned}
        L_{ops}&=-\frac{1}{T}\sum_{i=1}^T y_i\log \hat{y}_{ops}^i\\
        \hat{y}_{ops}^i&=\mathrm{SoftMax}(\mathrm{FC}(Z_{ops}))
    \end{aligned}
\end{equation}
where $T$ is the batch size, $\mathrm{FC(\cdot)}$ is the fully connected layer.

Additionally, to  improve intra-class compactness and inter-class separability, we incorporate a contrastive loss $L_{con}$. Given a batch of apps, the loss function is defined as follow:
\begin{equation}
\begin{aligned}
    L_{con} = &\frac{1}{|P|} \sum_{(i,j) \in P} d(Z_{ops}^i,Z_{ops}^j) + \\
&\frac{1}{|N|} \sum_{(i,j) \in N} \max\left(0, m - d(Z_{ops}^i,Z_{ops}^j)\right)
\end{aligned}
\end{equation}
$d(Z_{ops}^i,Z_{ops}^j)=\sqrt{(Z_{ops}^i-Z_{ops}^j)^\top \Lambda(Z_{ops}^i-Z_{ops}^j)}$ is the Mahalanobis distance. $Z_{ops}^i$ and $Z_{ops}^j$ refer to two cross-modal representations. $\Lambda$ is a positive semi-definite matrix. $P$ denotes the number of all positive sample pairs, $N$ denotes the number of all negative sample pairs, and $m$ is a threshold. Follow the principle of the contrastive learning, given a sample $Z_{ops}^i$, all other samples $Z_{ops}^j$ with the same label of $Z_{ops}^i$ are considered positive pairs, and those with different labels are negative pairs. 
\section{Experiment}
To evaluate the performance of our method, we conducted a comparative analysis of BIDO against other typical baselines. This evaluation is designed to examine the following seven research questions:

\textbf{RQ1}: How robust is BIDO under different OOD scenarios?

\textbf{RQ2}: What is the interpretability of BIDO?

\textbf{RQ3}: Which modules most significantly influence BIDO's performance?

\textbf{RQ4}: What are the effects of hyperparameters on BIDO's performance?

\subsection{Baselines}
We selected baselines based on the following criteria: (1) Impact: These models are outstanding research results in malware detection; (2) Diversity: We hope the selected baselines fall into different categories; (3) Reproducibility: To ensure fair comparison, we only consider detectors that provide source code. Accordingly, we select six baselines, three image-based (DexRay, Dex-CNN, and MADRF-CNN), two string-based (XMal and DetectBERT), and one graph-based (Malscan) approach.
Since our work focuses on improving robustness and interpretability within the static detection paradigm, dynamic approaches are not included as baselines in this study.
The details of the baselines are as follows:
\begin{itemize}
    \item \textbf{DexRay} \cite{daoudi2021dexray}: The model transforms the bytecode of DEX files into a grayscale vector image, and leverages a 1D CNN model for classification.
    \item \textbf{Dex-CNN} \cite{xiao2019image}: Instead of using grayscale images, Dex-CNN transforms DEX files into RGB images and applies a CNN for malware detection.
    \item \textbf{MADRF-CNN} \cite{zhu2023effective}: To enhance robustness against code obfuscation and concept drift, MADRF-CNN discardes the header and data sections of the DEX file and employes diverse pooling kernels to select informative subregions.
    \item \textbf{XMal} \cite{wu2021android}: The model utilizes API calls and permissions as features to construct a malware classifier. Unlike our approach, XMal does not use all available permissions but instead selects 158 based on a predefined list.
    \item \textbf{DetectBERT} \cite{sun2024detectbert}: It represents Smali codes of the APK as a class-level embedding and conducts malware classification based on it.
    \item \textbf{Malscan} \cite{wu2025malscan}: To obtain the coarse-grained malware representation, Malscan models the application's DFG as a social network and extracts structural features by analyzing the centrality of sensitive API nodes.
\end{itemize}

\subsection{Dataset}
As existing benchmarks cannot support evaluations on code obfuscation and concept drift simultaneously, we construct two datasets, Data-Ideal and Data-Obfu. The samples of these two dataset are downloded from the  Google Play \cite{googleplay} (the official Android app distribution platform) and two well researched benchmarks, Androzoo \cite{allix2016androzoo} and CICMalDroid2020 \cite{mahdavifar2020dynamic}. 
The Data-Ideal is constructed as follows:
\begin{itemize}
    \item Download APKs from Google Play, Androzoo and CICMalDroid2020.
    \item Assign labels for APKs. An app is labeled as benign if it is not detected by any antivirus from VirusTotal \cite{virustotal}, while an app is labeled as malicious if more than four detectors label it as malicious. As suggested by \cite{Zhu_Shi_Yang_Qin_Zhang_Song_Wang_2020}, the threshold is set to four due to the concern of label noises.
    \item Apps that do not contain DEX files or have abnormal formats are removed.
    \item Calibrate the balance rate of the dataset by the sampling technique.
\end{itemize}
Finally, Data-Ideal involves 12,375 malicious apps and 12,455 benign apps. 

To construct Data-Obfu, we implement six obfuscations on Data-Ideal. The details of each obfuscation method are as follows:
\begin{itemize}
    \item \textbf{ClassRename \& MethodRename} changes the names of classes or methods to arbitrary strings \cite{dong2018understanding}. 
    \item \textbf{ResStringEncryption} employs encryption to protect sensitive information, such as URLs and API keys stored in DEX files, or permissions and configurations in the XML files. 
    \item \textbf{Control Flow Obfuscation} alters the control flow of an app, making it harder to follow the logical clues.
    \item \textbf{NewAlignment} disrupts standard alignment patterns of code and data structures, confusing static analysis tools and reverse engineering efforts.
    \item \textbf{NewSignature} modifies method and class signatures, including names, parameter lists, and return types. 
    \item \textbf{Junk Code Insertion} injects irrelevant or meaningless code into the app’s source code without altering its functionality \cite{li2019obfusifier}.
\end{itemize}

According to \cite{Hammad_Garcia_Malek_2018}, combined obfuscation can cause more severe performance degradation. Based on this conclusion, we adopted the open-source tool, Obfuscapk \cite{aonzo2020obfuscapk} to implement these obfuscations sequentially. After removing the apps that encountered errors during obfuscation, the resulting Data-Obfu database coontains 12,088 malicious and 11,044 benign apps. 

\subsection{Evaluation Index}
We employ four commonly used metrics, Accuracy, Precision, Recall, and F1-Score, to evaluate the performance of our model and baselines. The specific definitions are as follows:
\begin{equation}
\begin{aligned}
    &{\rm Accuracy}=\frac{TP+TN}{TP+FP+TN+FN}\\
    &{\rm Precision}=\frac{TP}{TP+FP}\\
    &{\rm Recall}=\frac{TP}{TP+FN}\\
    &{\rm F1}=\frac{2\times ({\rm Precision}\times{\rm Recall})}{{\rm Precision}+{\rm Recall}}
\end{aligned}
\end{equation}
where TP (true positive) refers to the number of malicious applications accurately identified as malware. TN (true negative) indicates the number of benign applications correctly recognized as benign. FP (false positive) denotes the number of benign applications mistakenly classified as malware. FN (false negative) signifies the number of malicious applications incorrectly labeled as benign.

\subsection{Experiment Setting}

All experiments were conducted using an NVIDIA RTX 3090 GPU with 24 GB of RAM. 80\% of the dataset was used for training detectors, 10\% for validation, and the remaining 10\% for testing the performance of the proposed model and baselines. The number of the local feature map $K$ was set to 32. The number of epochs was set to 64, and the batch size was set to 8. Stochastic Gradient Descent (SGD) with momentum was used to update the parameters of the neural network model, with the momentum value set to 0.9. The initial learning rate was set to 0.001, and every two epochs, the learning rate is exponentially decayed by a factor of 0.9, which helps alleviate oscillations and instability during training. The loss weights for $\alpha$, $\beta$, $\pi$, $\gamma$, and $\delta$ were set to 0.1, 1.0, 0.005, 1.0, and 0.1, respectively.

\subsection{Results for RQ1}
To answer the first RQ, we design three scenarios. The first scenario is designed to test the performance of BIDO without OOD samples. The second and third scenarios are designed to test the robustness of BIDO against code obfuscation and concept drift respectively.

\subsubsection{The Results of BIDO without OOD Samples}
We divide the Data-Ideal dataset into training, validation, and testing sets with ratios of 80\%, 10\%, and 10\%.
The experimental results are presented in Table~\ref{tab:RQ1}, where bold in the table indicates the best performance and italics indicate the second best performance.

\begin{table}[h]
\centering
\caption{Results of Data-Ideal Dataset}
\label{tab:RQ1}
\begin{tabular}{lcccc}
\toprule
Method     & Accuracy & Precision & Recall & F1-score \\
\midrule
DexRay     & 91.46    & 94.57     & 88.90  & 91.65    \\
Dex-CNN    & 91.84    & 93.43     & \textit{90.72}  & 92.05    \\
MADRF-CNN  & 92.20    & \textit{94.84}     & 89.95  & 92.33    \\
XMal       & 88.24    & 87.99     & 86.77  & 87.38    \\
DetectBERT       & 90.57	&91.28	&88.94	&90.09   \\
Malscan    & \textit{93.50}    & \textbf{97.57}     & 89.17  & \textit{93.18}    \\
BIDO       & \textbf{94.42}    & 95.35     & \textbf{93.46}  & \textbf{94.48}\\
\bottomrule
\end{tabular}
\end{table}

As shown in Table~\ref{tab:RQ1}, the performance varies across different paradigms. 
Among all baselines, string-based methods exhibit the weakest performance. Their reliance on tokenized semantic features makes them incapable of capturing higher-order structural dependencies, resulting in limited performance. Image-based methods, including DexRay, Dex-CNN, and MADRF-CNN, achieve the second performance but remain inferior to the graph-based method. Notably, the two RGB image-based methods, Dex-CNN and MADRF-CNN, consistently outperform the greyscale-based DexRay, which partially confirms that RGB encoding retains richer semantic information than greyscale representations \cite{zhu2023effective}. However, all image-based baselines rely on discriminative CNNs that primarily capture global visual patterns. Their performances consistently lag behind the graph-based method and our approach, which aligns with recent findings that over-reliance on global features may degrade detection performance \cite{gao2024comprehensive}.
The graph-based method Malscan achieves the best performance among all baselines. By explicitly modeling structural dependencies, graph representations are able to capture fine-grained structural relationships, which provides better discriminative features than string or pixel-level image features.

In contrast, BIDO achieves the best performance across all evaluation metrics, including Accuracy, Recall, and F1-score. We believe that two techniques adopted in BIDO, the $1 \times 1$ convolutional mask and probabilistic space, contribute to this result. First, the function of the $1 \times 1$ convolutional mask is to identify informative subregions and reduce the dimension of the feature map. This contributes not only to computational efficiency but also to robust feature learning. Second, unlike graph-, image-, and string-based methods that embed samples as deterministic vectors in feature space, BIDO embeds each instance into a GMM latent space, which provides a clear and compact class-wise clustering structure, encoding both uncertainty and discriminative information in clusters. Consequently, BIDO demonstrates consistent performance gains over image-based baselines, exceeding DexRay, Dex-CNN, and MADRF-CNN by 2.83\%, 2.43\%, and 2.15\% in F1-score, respectively.

\begin{center}
\fcolorbox{black}{gray!10}{\parbox{.9\linewidth}{Although RGB-based image representations generally outperform grayscale ones, discriminative CNN-based methods that rely on global features are inherently limited in achieving optimal detection performance, even without OOD samples.
}}
\end{center}

\subsubsection{The Robustness of BIDO to Obfuscation}
To evaluate the robustness to code obfuscation, we conducted an experiment under three extreme  configurations. The details of configurations is presented in Table~\ref{tab:RQ2}.
\begin{table}[h]
\centering
\caption{Configurations for RQ2}
\label{tab:RQ2}
\begin{tabularx}{\columnwidth}{l X c c}
\toprule
No. & Training Data & Validation Data & Testing Data \\
\midrule
    1 & Data-Ideal (80\%) + Data-Obfu (20\%) & Data-Obfu & Data-Obfu \\
    2 & Data-Ideal (50\%) + Data-Obfu (50\%) & Data-Obfu & Data-Obfu \\
    3 & Data-Ideal (100\%) & Data-Obfu & Data-Obfu \\
\bottomrule
\end{tabularx}
\end{table}

\begin{table}[h]
	\centering
	\caption{Results of The First Configuration}
	\label{tab:2.1}
	\begin{tabular}{lcccc}
		\hline
		Method & Accuracy & Precision & Recall & F1-score \\
		\hline
		DexRay     & 87.67 & 88.97 & 86.00 & 87.46 \\
		Dex-CNN    & 87.37 & 88.68 & 85.67 & 87.15 \\
        MADRF-CNN  & 87.23 & 90.04 & 83.73 & 86.77 \\
		XMal       & 87.99 & 85.66 & \textbf{91.26} & 88.37 \\
		DetectBERT  & 86.85	&88.42	&83.76	&86.02 \\
		Malscan    & \textit{89.19} & \textbf{95.51} & 82.41 & \textit{88.48} \\
		BIDO       & \textbf{90.00} & \textit{91.67} & \textit{88.00} & \textbf{89.80} \\
		\hline
	\end{tabular}
\end{table}

\begin{table}[h]
\centering
\caption{Results of The Second Configuration}
\label{tab:2.2}
\begin{tabular}{lcccc}
\hline
Method & Accuracy & Precision & Recall & F1-score \\
\hline
DexRay     & 88.75 & 92.50 & 84.33 & 88.23 \\
Dex-CNN    & 89.90 & 88.64 & 91.53 & 90.06 \\
MADRF-CNN  & 90.60 & \textit{93.48} & 86.80 & 90.02 \\
XMal       & 88.69 & 85.45 & \textbf{93.26} & 89.18 \\
DetectBERT  & 87.62	&89.00	&85.75	&87.34 \\
Malscan    & \textit{90.92} & 91.02 & 90.95 & \textit{90.98} \\
BIDO       & \textbf{92.73} & \textbf{93.72} & \textit{91.60} & \textbf{92.65} \\
\hline
\end{tabular}
\end{table}

\begin{table}[h]
\centering
\caption{Results of The Third Configuration}
\label{tab:2.3}
\begin{tabular}{lcccc}
\hline
Method & Accuracy & Precision & Recall & F1-score \\
\hline
DexRay     & 69.50 & 75.35 & 61.61 & 67.79 \\
Dex-CNN    & 53.90 & 57.80 & 42.67 & 49.10 \\
MADRF-CNN  & \textit{74.20} & 84.04 & 62.32 & 71.57 \\
XMal       & 70.65 & 62.45 & \textbf{97.14} & \textit{76.02} \\
DetectBERT  & 57.60	&58.37	&56.19	&57.26 \\
Malscan    & 66.11 & \textbf{87.92} & 41.09 & 56.00 \\
BIDO       & \textbf{81.30} & \textit{84.84} & \textit{78.06} & \textbf{81.31} \\
\hline
\end{tabular}
\end{table}

The experimental results of different configurations are presented in Table~\ref{tab:2.1},~\ref{tab:2.2},~\ref{tab:2.3}, which demonstrate that code obfuscation degrades the performance of all detectors. Nevertheless, our approach consistently outperforms all baselines across three evaluation metrics.

Specifically, by comparing Table.~\ref{tab:RQ1} to Table~\ref{tab:2.1},~\ref{tab:2.2}, we can draw the following conclusions:

Image-based Methods (DexRay, Dex-CNN, and MADRF-CNN) suffer from the most significant performance degradation under obfuscated scenarios. The results align with the findings of Gao et al. \cite{gao2024comprehensive}, indicating that image-based approaches struggle to maintain robustness. We attribute this to the fact that image-based detectors heavily rely on the byte sequences of DEX files. Obfuscation techniques such as ClassRename or ResStringEncryption can easily change the visual pattern of DEX files. Notably, although MADRF-CNN attempts to mitigate the effect of obfuscation by discarding DEX headers and data sections, it still experiences the sharpest decline in accuracy. This suggests that simply removing irrelevant information is insufficient to counter the obfuscation.

In contrast, graph (Malscan) and string-based methods (XMal and DetectBERT) demonstrate relatively robust performance, which aligns with recent findings in \cite{gao2024comprehensive}. This result stems from the coarse-grained nature of their feature. Unlike fine-grained byte-level features, their high-level features, such as function call graphs and permission are intrinsically more difficult to obfuscate without compromising the application's core functionality. Furthermore, the moderate degradation suggests that raph-based and string-based detectors are particularly less affected by code-level obfuscation.

As illustrated in Table \ref{tab:2.3}, BIDO consistently achieves the best performance across all test scenarios, particularly in the most extreme cases (e.g., the training set contains no obfuscated samples). We believe that this superior robustness stems from the following two techniques: (1) Likelihood-based OOD Discrimination: In BIDO, every instance is assigned a likelihood. A lower likelihood indicates a higher probability that an instance is an OOD sample. By quantifying this likelihood, the model gains the ability to discriminate between familiar patterns and the "unknown" shifts introduced by OOD samples. This prevents the model from making overconfident but incorrect predictions on OOD samples. (2) GMM-based representation. In the GMM space, the representation of each instance is correlated only to two statistical parameters: the mean and variance of the Gaussian distribution. By focusing on these statistical invariants rather than fine-grained feature, the model effectively abstracts away the noise generated by obfuscation. This parameter-based representation ensures that the latent features provide a more robust foundation for classification.

\begin{center}
\fcolorbox{black}{gray!10}{\parbox{.9\linewidth}{Obfuscation substantially degrades the performance of all detectors. GMM-based latent space can boost the robustness of representation.}}
\end{center}

\subsubsection{The Robustness of BIDO to Concept Drift}
The training and validation data set of this RQ are selected from the samples of 2016 in Data-Ideal, and the testing dataset is selected from the samples of 2017 in Data-Ideal. In total, the dataset includes 1250 malicious and 1250 benign apps.
\begin{table}[h]
\centering
\caption{Results of Concept Drift}
\begin{tabular}{lcccc}
\toprule
Method & accuracy & precision & recall & F1-score \\
\midrule
DexRay & 74.50 & \textit{78.16} & 68.00 & 72.73 \\
Dex-CNN & 64.33 & 69.46 & 56.56 & 62.35 \\
MADRF-CNN & 75.80 & 69.49 & \textbf{92.00} & 79.17 \\
XMal & \textit{82.14} & 77.28 & \textit{90.83} & \textit{83.51} \\
DetectBERT  & 79.33	&78.50	&82.14	&80.26 \\
Malscan & 81.45 & 76.56 & 87.89 & 81.45 \\
BIDO & \textbf{86.00} & \textbf{86.89} & 84.80 & \textbf{85.83} \\
\bottomrule
\end{tabular}
\label{tab:concept_drift}
\end{table}

As shown in Table~\ref{tab:concept_drift}, compared to code obfuscation, concept drift leads to more severe performance degradation across all detectors. Despite this, BIDO achieves the best overall performance with an F1-score of 85.83\%, significantly surpassing all baselines.

Specifically, string-based detectors XMal achieves the second-best F1-score and the second-highest recall. The graph-based method Malscan also exhibits competitive performance, with an F1-score of 81.45\%. These results align with prior findings \cite{gao2024comprehensive}, confirming that features derived from high-level behavioral abstractions, such as permissions, Data Flow Graphs (DFG), and API call sequences, evolve more slowly. This allows them to remain effective even as the underlying raw code undergoes frequent updates.
Image-based detectors (DexRay, Dex-CNN, and MADRF-CNN) suffer more severe performance degradation. Since image-based methods map raw bytecode directly into visual patterns, they are more sensitive to the frequent Android app evolution such as system upgrades, bug fixes, or feature enhancements \cite{gao2024comprehensive}. Even minor bytecode modifications can result in "feature shifts" in the image space that the models fail to recognize.

The superior performance of BIDO under concept drift further validates the generalization of the generative classification framework. BIDO treats concept drifted samples as potential OOD instances. This allows the model to quantify the "drift" rather than making erroneous classifications based on outdated features. Additionally, the representaiton generated by BIDO focus on statistical invariants that the core representation remains stable despite the noise introduced by app evolution.

\begin{center}
\fcolorbox{black}{gray!10}{\parbox{.9\linewidth}{Concept drift leads to more severe performance degradation than code obfuscation. BIDO achieves better resistance to concept drift.}}
\end{center}

\subsection{Results for RQ2}
In this section, we demonstrate the reliability of the interpretation generated by BIDO. According to the Section.~\ref{GC}, each latent representation $Z_{lat}$ corresponds to a Gaussian centroid $u_{y_j}$ and a class-conditional likelihood $p(Z_{lat}^i|y_j)=\mathcal{N}\left(Z_{lat}^i, u_{y_j},1\right)$. So, the relative distances of latent representation to the centers of different classes of malware  $||Z_{lat}^i-u_{y_j}||_2^2$ can be used to interpret their semantical similarities. Further more, the class-conditional likelihood $p(Z_{lat}^i|y_j)$ can be used to quantify the confidence of the identified semantical similarity. 

This explanation is unique to our generative classification and it is impossible for discriminative classification models to achieve such explanation. The discriminative classification model focus only on learning features or logits to separate classes. There is no latent space in which the input data is embedded in a way that preserves explicit probabilistic geometry. So, it is impossible for discriminative classification models to interpret "\textit{how well this image matches class A vs B}".

\begin{figure}[tb]
	\centering
	\begin{subfigure}[b]{0.24\textwidth}
		\centering
		\includegraphics[width=\linewidth]{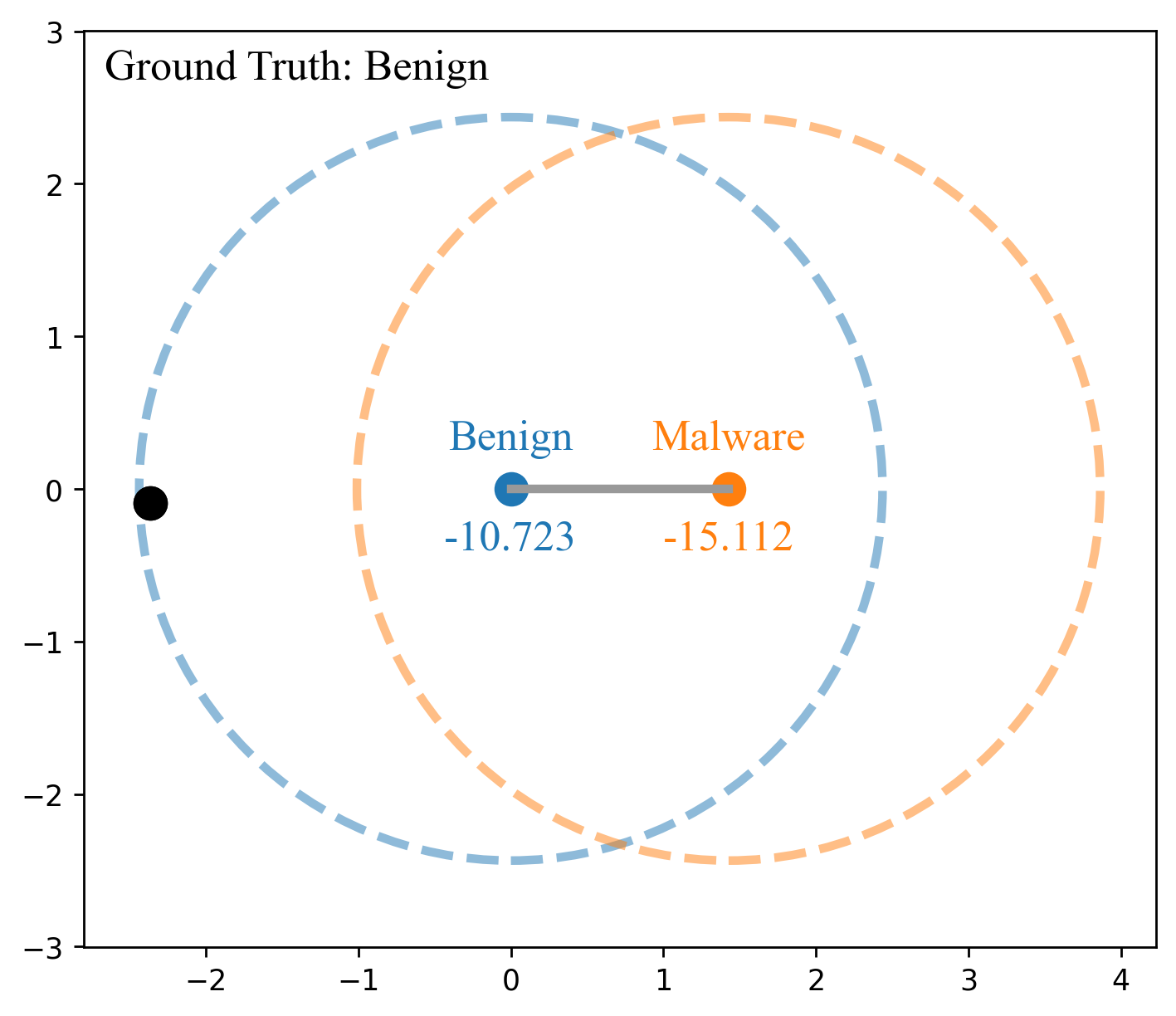}
		\caption{Bengin Sample}
		\label{fig:sub1}
	\end{subfigure}
	\hfill
	\begin{subfigure}[b]{0.24\textwidth}
		\centering
		\includegraphics[width=\linewidth]{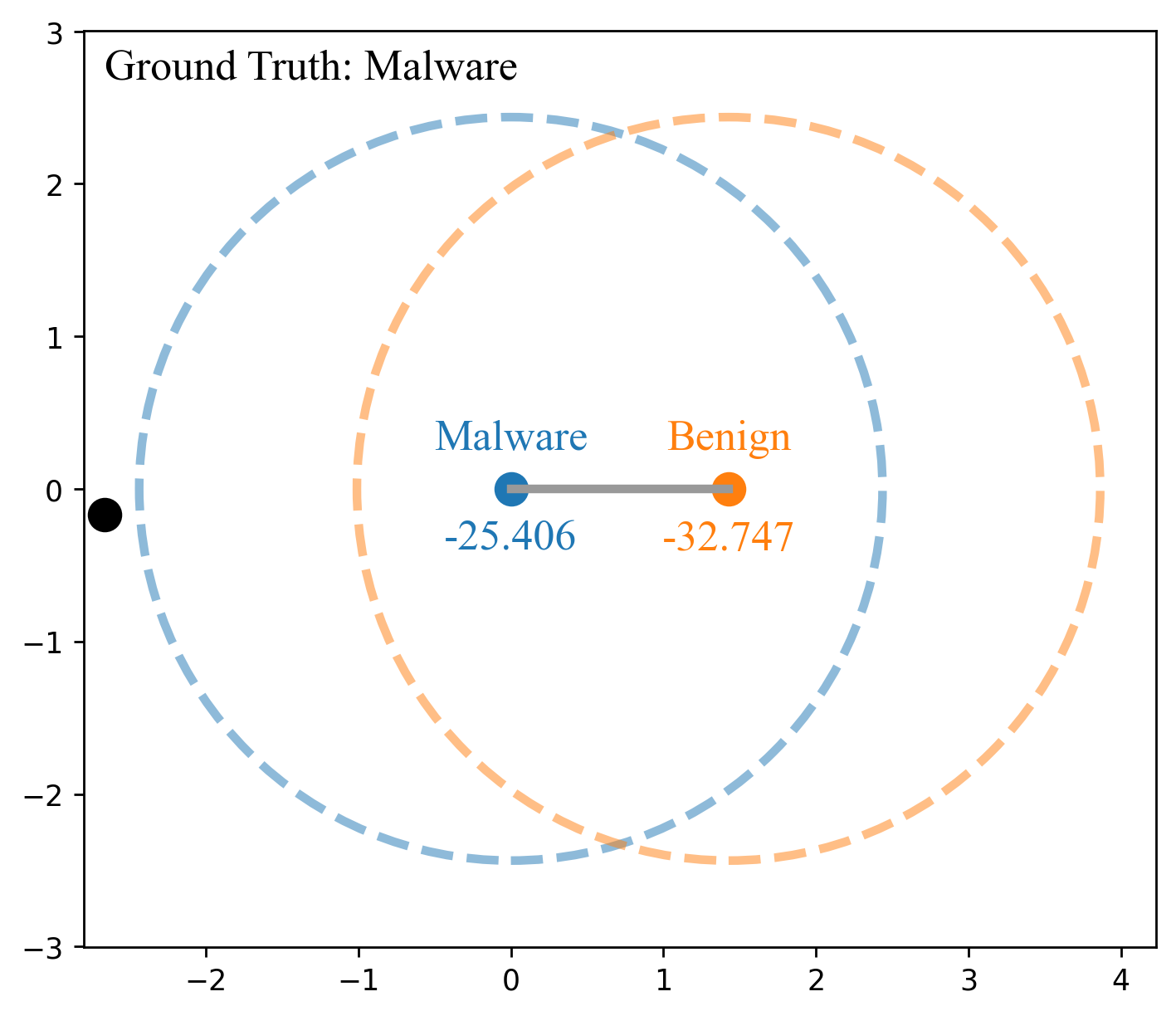}
		\caption{Malware Sample}
		\label{fig:sub2}
	\end{subfigure}
	\vspace{0.5em}
	\begin{subfigure}[b]{0.24\textwidth}
		\centering
		\includegraphics[width=\linewidth]{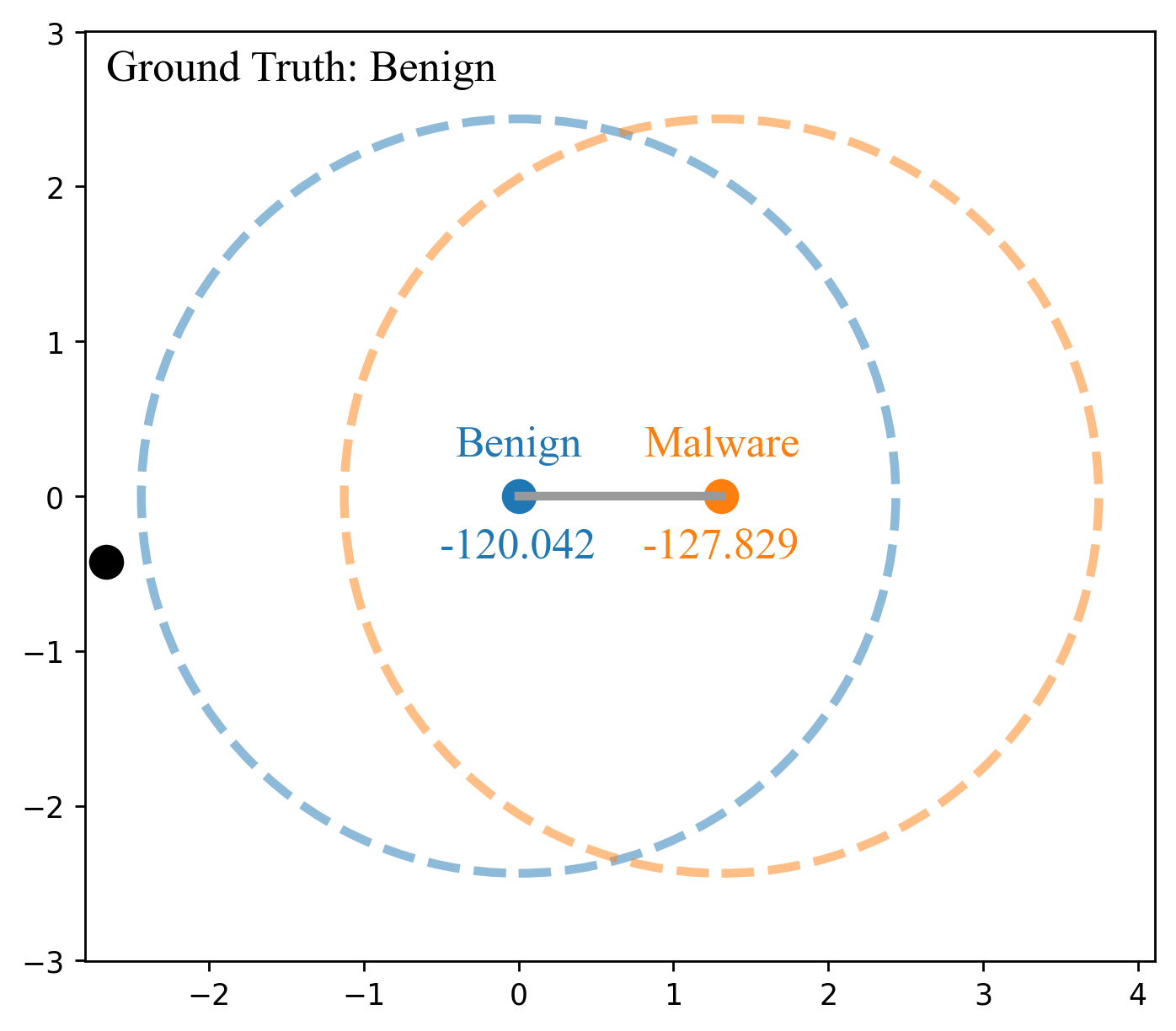}
		\caption{Obfuscation Sample}
		\label{fig:sub3}
	\end{subfigure}
	\hfill
	\begin{subfigure}[b]{0.24\textwidth}
		\centering
		\includegraphics[width=\linewidth]{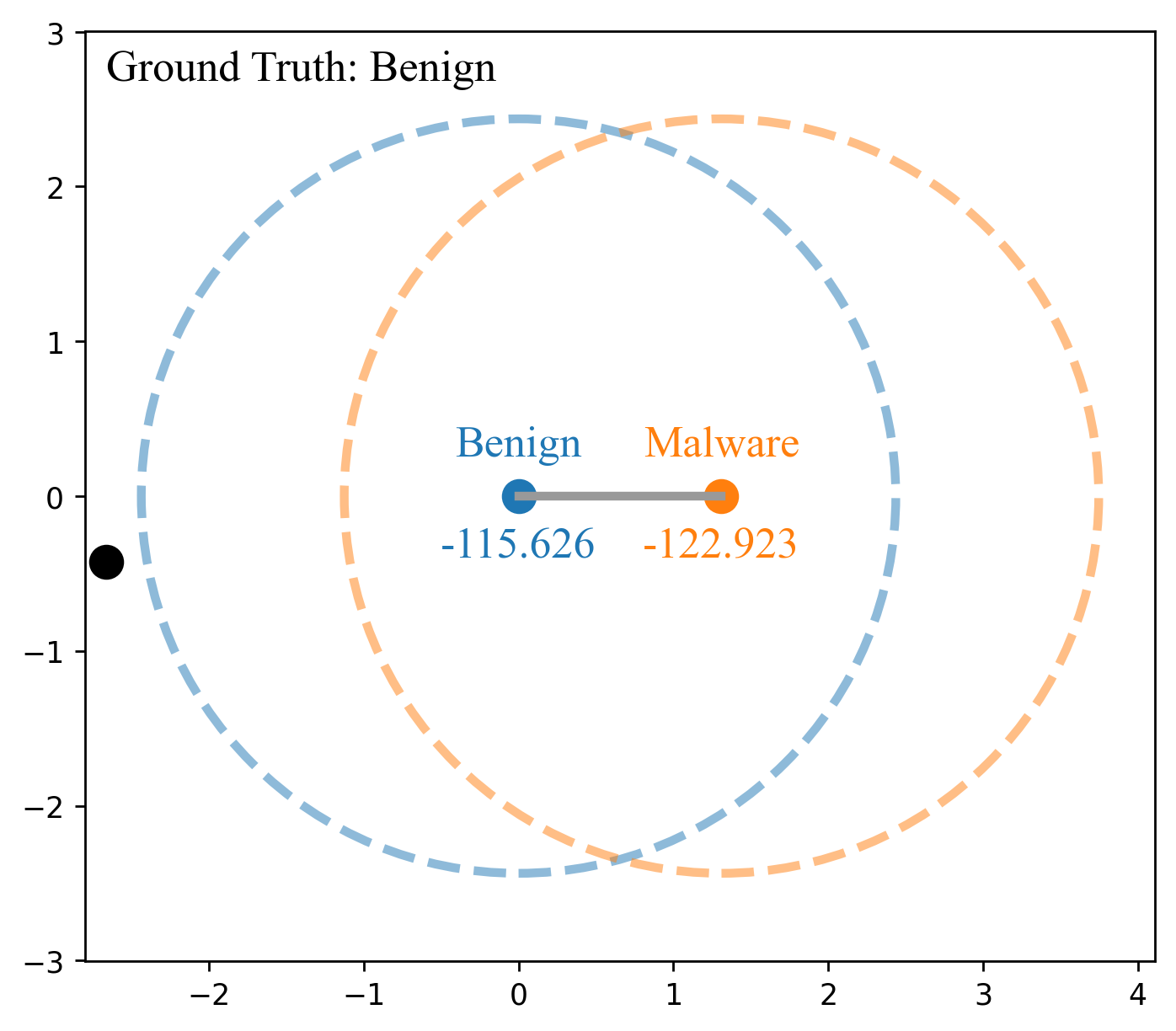}
		\caption{Concept Drift Sample}
		\label{fig:sub4}
	\end{subfigure}
	\caption{Latent representations of input samples (black point) and the centrids of malware (orange point) and beign (blue point) classes. }
	\label{fig:interpretation}
\end{figure}

To verify our belief, we represent the latent representation of input impages and the malware and bengin Gaussian distributions into two dimentional space. The circles represent 90\% of the mass of each Gaussian distribution. The values affiliated to the centriods are the likelihoods of the input sample belongs to the distribution. 
We could see from the Fig.~\ref{fig:interpretation} that the malware class has large overlap with bengin class. It means samples lying in the overlap zone are far more difficult to classify than the sample locates out of the overlap zone. The detection results of OOD samples are less confident than that of ordinary samples. Furthermore, samples corresponding to code obfuscation and concept drift exhibit significantly lower likelihood values compared to ordinary samples. This observation suggests that the model holds lower confidence in its detection results for these inputs.

\begin{center}
	\fcolorbox{black}{gray!10}{\parbox{.9\linewidth}{The distance between the latent representation to the centroids of the malware and benign classes serves as a criterion for interpretating its class membership. Likelihood values further provide a quantitative measure of confidence in the detection results. }}
\end{center}

\subsection{Results for RQ3}
To identify the effects of different modules of BIDO, we conduct an ablation experiment on the Data-Ideal dataset. The variants are defined as follows:

\begin{itemize}
    \item M-xml: Denote the BIDO after removing the DEX-image generation, local feature selection,cross-modal dependency and generative classification modules. The detection results are generated by a MLP layer.
    \item M-dex: Denote the BIDO after removing the XML-image generation and cross-modal dependency and generative classification modules. The detection results are generated by a MLP layer.
    \item M-gc. The only difference between M-gc with M-dex is the  detection results are generated by generative classification module.
    \item M-fusion. Denote the BIDO after removing the generative classification module. The detection results are generated by a MLP layer.
\end{itemize}

\begin{table}[h]
\centering
\caption{The Impacts of Different Components on Detection Results}
\begin{tabular}{lllll}
\hline
Method    & accuracy       & Precision      & recall         & F1-score       \\ \hline
M-xml       & 89.21          & 91.17          & 87.69          & 89.40          \\
M-dex       & 92.64          & 95.40          & 90.15          & 92.70          \\
M-gc        & 93.75          & \textbf{96.44} & 91.32          & 93.81          \\
M-fusion    & \textit{94.15} & 94.17          & \textbf{94.23} & \textit{94.20} \\
BIDO & \textbf{94.42} & \textit{95.53} & \textit{93.46} & \textbf{94.48} \\ \hline
\label{tab:RQ4}
\end{tabular}
\end{table}

According to results presented in Table~\ref{tab:RQ4}, we can observe: First, M-gc achieves the highest Precision. This suggests that modeling the underlying data distribution significantly enhances precision. Second, by fusing XML and DEX features, M-fusion achieves the highest Recall, outperforming single-modality variants. This indicates that integrating cross-modal information captures more diverse malware patterns, reducing false negatives. Finally, the full BIDO model integrates these complementary strengths, precision from generative modeling and recall from cross-modal feature fusion, to achieve the best overall performance, with the highest Accuracy and F1-score.

\begin{center}
\fcolorbox{black}{gray!10}{\parbox{.9\linewidth}{ Generative classification can significantly improve the precision and cross-modual representation can boost the recall. These two modules are complementary, and their combination can significantly enhance detector's performance.}}
\end{center}

\subsection{Results for RQ4}
The number of local feature maps, denoted as $K$, is a crucial hyperparameter in our approach. To assess its impact, we conducted a comparative study by setting different values across 2, 4, 8, 16, 32, and 64. According to the results shown in Table VIII, we can see that performance improves steadily as $P$ increases. Specifically, when $K = 2$, the accuracy, precision, recall, and F1-score are  85.94\%, 86.69\%, 84.66\%, and 85.66\%, respectively. As $K$ increases from 4 to 32, all metrics show continuous improvement. However, further increasing $K$ to 64 does not yield additional gains. Therefore, $K = 32$ appears to be the optimal setting for our method. The experiments for RQ6 were conducted on the Data-Ideal dataset.

\begin{table}[!htb]
\centering
\label{tab:RQ6}
\caption{The results of our method under different numbers of local feature maps}
\begin{tabular}{ccccc}
\hline  $\textbf{\textit{K}}$ & \textbf { Accuracy } & \textbf { Precision } & \textbf { Recall } & \textbf { F1-score } \\
\hline 
$K$=2 & 85.94 & 86.69 & 84.66 & 85.66 \\
$K$=4 & 84.38 & 86.36 & 85.41 & 85.88 \\
$K$=8 & 87.46 & 87.52 & 87.72 & 87.62 \\
$K$=16 & 89.69 & 91.67 & 92.98 & 92.32 \\
$K$=32 & \textbf{94.42} & \textbf{95.35} & \textit{93.46} & \textbf{94.48} \\
$K$=64 & \textit{91.29} & \textit{92.38} & \textbf{94.74} & \textit{93.54} \\
\hline
\end{tabular}
\end{table}

\begin{center}
\fcolorbox{black}{gray!10}{\parbox{.9\linewidth}{Increasing the number of local feature maps can improve the model’s performance, but the optimal configuration requires manual tuning.}}
\end{center}

\section{Conclusion}
Although numerous image-based malware detectors have been proposed in recent years, only a limited number demonstrate strong resistance to code obfuscation and concept drift. Unlike most existing approaches, which treat code obfuscation and concept drift as separate challenges, we propose a unified solution that addresses both issues from the common statistical root, OOD. Experimental results not only demonstrate that our method significantly outperforms all baselines, but also reveal several findings that could guide future research:
\begin{itemize}
    \item Generative classification enhances robustness against OOD samples by modelling the inherent distribution of malware, where conventional discriminative detectors fail. Moreover, generative models substantially improve the reliability of interpretation, as interpretations are produced concurrently with the detection process.
    \item Concept drift results in more severe performance degradation than code obfuscation.
    \item Features from DEX files and configuration file are complementary and their fusion can enhance the performance of malware detectors.
\end{itemize}

\ifCLASSOPTIONcaptionsoff
  \newpage
\fi



%


\bibliographystyle{IEEEtran}
\bibliography{references}

%








\end{document}